\documentclass[conference]{IEEEtran}
\IEEEoverridecommandlockouts
\usepackage{cite}
\usepackage{amsmath,amssymb,amsfonts}
\usepackage{algorithmic}
\usepackage{graphicx, subcaption}
\usepackage{textcomp}
\usepackage{xcolor, xfrac, braket, nicefrac, hyperref}
\usepackage{qcircuit}
\def\BibTeX{{\rm B\kern-.05em{\sc i\kern-.025em b}\kern-.08em
    T\kern-.1667em\lower.7ex\hbox{E}\kern-.125emX}}
\begin{document}

\title{A Quantum-Classical Hybrid Method for Image Classification and Segmentation}

\author{\IEEEauthorblockN{Sayantan Pramanik$^\dagger$, M Girish Chandra$^\ast$, C V Sridhar$^\dagger$,  Aniket Kulkarni$^\dagger$, Prabin Sahoo$^+$, Vishwa Chethan D V$^\dagger$, \\ Hrishikesh Sharma$^\ast$, Ashutosh Paliwal$^\dagger$,  Vidyut Navelkar$^\dagger$, Sudhakara Poojary$^\dagger$, Pranav Shah$^\dagger$, Manoj Nambiar$^\ast$}
\IEEEauthorblockA{\textit{$^\dagger$TCS Incubation \hspace{0.2cm} $^\ast$TCS Research \hspace{0.2cm} $^+$TCS Manufacturing and Utilities} \\
\{sayantan.pramanik, m.gchandra, sridhar.cv, aniket.k, prabin.sahoo, v.dv2, hrishikesh.sharma, ashutosh.paliwal, \\  vidyut.navelkar, sudhakara.poojary, pranav.shah, m.nambiar\}@tcs.com}
}

\maketitle

\begin{abstract}
Enormous activity in the Quantum Computing area has resulted in considering them to solve different difficult problems, including those of applied nature, together with classical computers. An attempt is made in this work to nail down a pipeline consisting of both quantum and classical processing blocks for the task of image classification and segmentation in a systematic fashion. Its efficacy and utility are brought out by applying it to Surface Crack segmentation. Being a sophisticated software engineering task, the functionalities are orchestrated through our in-house Cognitive Model Management framework.
\end{abstract}

\begin{IEEEkeywords}
quantum computing, variational quantum classifiers, q-means, data encoding, ansatz, measurement, quantum software development life cycle, computer vision, image classification, image segmentation, crack detection
\end{IEEEkeywords}

\section{Introduction}
Image Segmentation and Classification is an extensively researched area with applications spanning across domains. Because of its importance, many classical approaches exist. With the emerging scenario of Quantum Computing, there is thrust to explore them in these tasks, either for a possible speed up or better some of the aspects, like, using a smaller number of training examples, more accuracy etc. One more aspect which is getting settled is the fact that even when improved quality and good enough sized quantum computers are available in the future, both quantum and classical computing work and “cooperate” together to solve useful real-life problems. Keeping this hybrid-architecture in mind, this paper attempts to suggest a possible pipeline of classical and quantum processing blocks to achieve image segmentation and further classification of these segments. In particular, we consider the classification and segmentation of Kaggle Surface Crack Data and in the process, suggest a rather generic pipeline to accomplish the task, taking into account the limitations of number of qubits and depth of the quantum circuits at present into account. The results obtained confirm the expected functionality of the proposal accompanied with quantitative metrics reinforcing its usefulness. Further, with different processing models (machine learning as well as others) of both quantum and classical variety working together, the software aspects involved in putting the pipeline into action are definitely complex. In order to negotiate this aspect and ``productize" this software solution in the future, our exclusive Cognitive Model Management (CMM) framework and OpSense tool are utilized to facilitate a systematic execution of the whole workflow.

In Section \ref{sec:overall}, the big picture of the pipeline is briefed; Section \ref{sec:dataset} touches upon the data set considered; Section \ref{sec:detailed} covers the details of various models in considerable detail, including segmentation, classification and optimization. Before providing comparative results, fully classical approaches are mentioned in Section \ref{sec:classical}, followed by Section \ref{sec:results} on results. There is a small section on Future Tasks (Section \ref{sec:future}). Section \ref{sec:qse} brings out our initiation into the Quantum Software Engineering pivoted on CMM and OpSense.

\section{Overall Pipeline}\label{sec:overall}
\begin{figure}
	\hspace{-0.28cm}
	\includegraphics[scale=0.34]{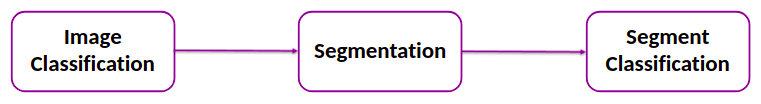}
	\caption{Generic pipeline for segmenting cracks in images, which involves the three steps highlighted in the simple flowchart.}
	\label{fig:overall}
\end{figure}

As shown in Fig. \ref{fig:overall}, the problem of crack detection and segmentation has been addressed in three steps:
\begin{enumerate}
	\item The first step involves classifying the images to detect whether it contains a crack or not. If it is found not to contain any anomalies, then the subsequent pipeline may be terminated for that particular input image.
	\item After classification, the images which are identified to have cracks in them are segmented into two different clusters. Depending on the size and resolution of the images, this could prove to be a time-consuming and computationally-intensive step, which need not be unnecessarily wasted on images which do not have any cracks. Hence the requirement of the first classification step. Further, clustering being an unsupervised algorithm, is expected to make the pipeline generic and transferable to other data sets for similar problems, without the requirement of retraining.
	\item Once the clusters have been identified in an image, the different regions can be classified to check which of them actually correspond to cracks.
\end{enumerate}

Different parts of the pipeline have been expounded upon in greater detail, along with more technical arguments, implementation steps and results, and subtleties in Sec. \ref{sec:detailed}.

\section{The Dataset}\label{sec:dataset}
\begin{figure}[h]
	\centering
	\begin{subfigure}[t]{0.24\textwidth}  
		\centering 
		\includegraphics[scale=0.5]{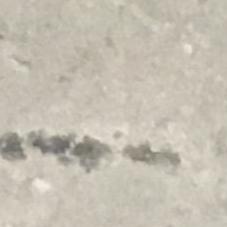}
		\caption{An image without any cracks in it, labelled "Negative"}
		\label{fig:negative}
	\end{subfigure}
	\begin{subfigure}[t]{0.24\textwidth}  
		\centering 
		\includegraphics[scale=0.5]{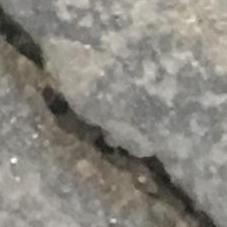}
		\caption{An image with a crack}
		\label{fig:positive}
	\end{subfigure}
	\caption{Examples of images with and without cracks.}
	\label{fig:example}
\end{figure}

The data set in question, namely the Surface Crack data set, is a set of $40,000$ low-resolution images that are openly available-to-all via Kaggle \cite{kaggle}. Each image is $227\times 227$, 3-channel RGB, accompanied with a binary label which identifies whether it contains a crack or not. The entire data set has been evenly-balanced between images with and without cracks, available as two folders named "Positive" and "Negative" with $20,000$ images in each. It must be noted that the images that fall under the "Negative" label do however contain dents, discolouration, and other forms of aberrations which are sometimes difficult to differentiate from actual cracks. Other surface patterns and textures exist in instances from both classes. Representative images, one from each folder, can be found in Fig. \ref{fig:example}.

\section{Detailed Pipeline and Intermittent Results}\label{sec:detailed}
\begin{figure*}
	\centering
	\includegraphics[scale=0.5]{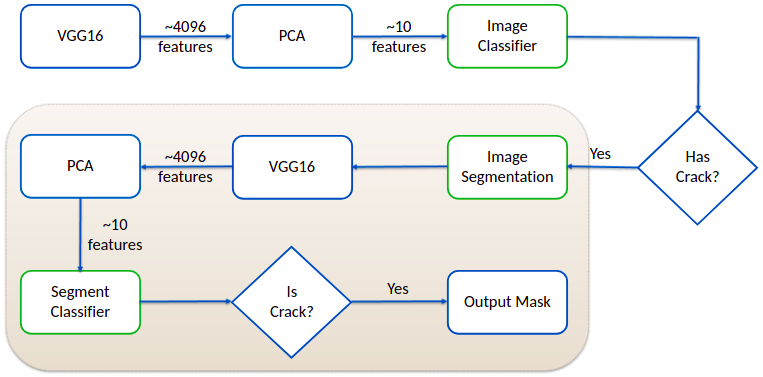}
	\caption{Detailed pipeline to solve the problem of crack segmentation in images. The green-coloured boxes are steps for which quantum algorithms have been implemented and utilized. The steps in the shaded box correspond to image segmentation and classifying the detected regions.}
	\label{fig:detailed}
\end{figure*}

An bried overview of the pipeline was presented in Sec. \ref{sec:overall}, the components of which have been dissociated into more granular steps in Fig. \ref{fig:detailed}. The process starts with the image classifier, for which features are extracted from the images using a pre-trained deep learning architecture (such as VGG16 \cite{VGG16} trained on the ImageNet data set, and stripped of its final classification layer). Considering the simplicity of the data set this approach is being tested on, it was concluded that using more complicated feature-extractors would be overkill. The VGG16 model extracts $4096$ features from each image which require too many qubits to encode into a quantum circuit using angle-embedding, or too deep circuits if amplitude-embedding is used instead. Better classical-to-quantum data encoding techniques are presently being researched \cite{encoding} to find an appropriate trade-off between circuit-depth and qubit-requirement. 

Keeping the limitations of presently-available simulators and quantum processors in mind, the dimensionality of the features obtained from VGG16 was reduced from $4096$ to the order of $10$. Specifically, within the scope of this work, Principal Component Analysis (PCA) has been used to reduce the $4096$ features to just $4$, which were then passed on to the first quantum image classifier for training and inference purposes. This is similar to the Classical+Quantum approach proposed in \cite{Schuld}. Alternative approaches that use quantum computing for feature extraction and classification have also been proposed \cite{Schuld, QCNN, Quanvolutional} that have been explored, but not implemented within the context of this work. Further, quantum classification can also be achieved through algorithms other than variational quantum classifiers, such as quantum versions of support-vector machines (QSVM) \cite{QSVM} post the feature-extraction stage.

Following image classification, the task of image segmentation has been attempted through a quantum counterpart of the $k$-means algorithm, known as $q$-means \cite{q-means}. Additional methods of image clustering such as graph-cuts \cite{graph-cuts} which model the pixels of an image as vertices of a graph and edges formed by pixel intensities of $n$-nearest neighbours, and image segmentation by Quantum Hadamard Edge Detection (QHED) \cite{QHED} also serve as interesting and viable approaches. As already discussed, all of the individual regions detected through image segmentation are passed on for another round of classification which uses the same techniques described for the aforementioned image classifier. The regions identified as cracks are highlighted as such in the original image, which concludes the process of crack segmentation \footnote{All experiments portrayed in this paper were developed on AWS Braket SDK, simulated on Braket's Local Simulator running on a generic personal computer with an Intel 10$^{th}$ generation i5-10310U processor with integrated graphics and 16GB of RAM. All quantum circuit diagrams were generated using Q-circuit \cite{qcircuit}.}.

\subsection{Segmentation by Quantized $k$-means}
Given a set of data points, each consisting of $d$-dimensional feature vectors, it is often necessary to segregate them into $k$ different baskets, The $i^{th}$ data point is given by $v_i = \{v_i^1, v_i^2, \dots, v_i^d\}$, $ i\in[N]$; and the baskets are characterized by cluster centroids, which themselves are d-dimensional vectors, $c_j=\{c_j^1, c_j^2, \dots, c_j^d\}$, $j\in[k]$. One starts with arbitrary guesses of the $k$ centroid vectors and the pairwise absolute distance between each cluster centroid and data point is calculated: $|d(v_i,c_j)|$. The $i^{th}$ data vector is assigned to the cluster to which it is situated closest. Having allocated each data point to a cluster, the centroids themselves are updated by taking the average of corresponding features of each vector in the cluster. The process is repeated to convergence to obtain the final clusters and centroids.

\subsubsection{$q$-means}
A quantum equivalent of the $k$-means algorithm, aptly dubbed $q$-means, has been proposed in \cite{q-means} which performs in poly-logarithmic complexity with respect to $N$, compared to being linear in $N$ for the classical implementation. The algorithm, though, comes with the caveat the it employs the use of a $QRAM$ \cite{QRAM} to achieve the said improvement in performance. $QRAM$ has been an area of active research over the recent years, but is not something that is available for immediate use. In this paper, we have attempted to circumvent this problem by using a somewhat naive strategy which works as a placeholder in the NISQ-era \cite{NISQ} until such memory units are available. Until such time, although a speed-up over classical techniques is not apparent, the exercise serves as a proof-of-principle which will be ready to provide benefit in the near future.

\subsubsection{Distance Metrics}
The cornerstone operation in classical $k$-means and quantum $q$-means algorithm is the calculation of distances between cluster centroids and the data vectors. The most common distance-metric of choice is the Euclidean distance $D_E$ between two points in $d$-dimensional space. In the quantum-regime, Euclidean distance calculation is done by expressing the data vectors and centroids in the form of statevectors, and an affine transform of the modulus of inner products between them give the requisite distance values:
\begin{equation} \label{eq:dist}
\begin{split}
D_E^2(\ket{x}, \ket{y}) &= |(\ket{x} - \ket{y})|^2 = (\bra{x} - \bra{y})(\ket{x} - \ket{y})\\
&= 2 - \braket{x|y} - \braket{y|x}\\
&= 2 - 2\braket{x|y}
\end{split}
\end{equation}
where the entries of $\ket{x}$ and $\ket{y}$ being real-valued, $\braket{x|y} = \braket{y|x}$. $|D_E|$ and $D_E^2$ having the same monotonicity, the later can be used as a way to quantify the distance between data and centroid vectors without adversely affecting the performance of the algorithm \cite{IonQ}. 

In the subsequent subsections, the details of some quantum algorithms that help calculate $D_E^2$, along with their implementation details, advantages and disadvantages, have been spelt.

\subsubsection{Swap Test}
\begin{figure}[!h]
\[
\Qcircuit @C=2em @R=1em @!R
{\lstick{\ket{0}} & \gate{H} & \ctrl{1} & \gate{H} & \meter & \cw \\
	\lstick{\ket{0}} & \gate{E(x)} & \qswap & \qw & \qw & \qw  \\
	\lstick{\ket{0}} & \gate{E(y)} & \qswap \qwx & \qw & \qw & \qw} \]
\caption{Quantum Circuit to implement Swap Test.}
\label{fig:swap}
\end{figure}
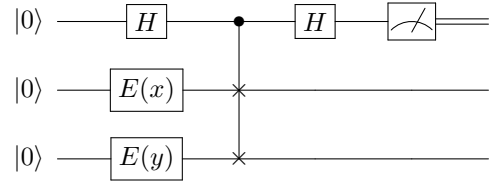

The Swap Test \cite{swap} is perhaps the most famous method of quantifying the overlap between two quantum states given by $\ket{x}$ and $\ket{y}$. The respective states are first prepared from $\ket{0}$ by the application of appropriate parametrized unitaries $E(x)$ and $E(y)$, respectively, which encode the points $x$ and $y$ into quantum states. The circuit in Fig. \ref{fig:swap} gives an overview of the overall procedure.

The probability of measuring the first qubit in the state $0$ in the computational basis is given by Eq. \eqref{eq:swap}, can be used to determine the value of $\braket{x|y}$, which can be retrofitted into Eq. \eqref{eq:dist} to get $D_E^2$.
\begin{equation} \label{eq:swap}
P(0) = \frac{1}{2} \left(1+\lvert\braket{x|y}\rvert^2 \right)
\end{equation}
However, the problem with using the Swap Test for determining an equivalent of Euclidean distance is multifold. The most obvious one is that Eq. \eqref{eq:swap} gives the value of $\lvert \braket{x|y}\rvert^2$, whereas Eq. \eqref{eq:dist} requires $\braket{x|y}$ to be evaluated. Further, the algorithm uses a controlled-swapping operation, which can prohibitively increase the depth of the circuit if the registers for $\ket{x}$ and $\ket{y}$ use multiple qubits.

\subsubsection{Hadamard Test}
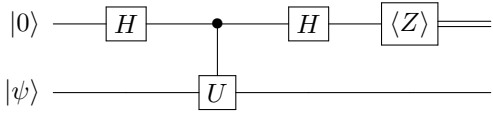
\begin{figure}[h]
	\[ \Qcircuit @C=2em @R=1em @!R{
		\lstick{\ket{0}} & \gate{H} & \ctrl{1} & \gate{H} & \gate{\braket{Z}} & \cw \\
		\lstick{\ket{\psi}} & \qw & \gate{U} & \qw
		& \qw & \qw }
	\]
	\caption{Quantum circuit to calculate $Re\left(\braket{\psi|U|\psi}\right)$ using Hadamard Test.}
	\label{fig:hadamard}
\end{figure}

Some of the difficulties faced in employing the Swap Test can be addressed through the use of Hadamard Test \cite{hadamard}, instead, the circuit depiction of which can be found in Fig. \ref{fig:hadamard}. The latter also uses only about half the qubits compared to the former. A quick and simple analysis of the algorithm suggests that the $Z$ expectation value of the first qubit is:

\begin{equation}
\braket{Z}_0 = Re\left(\braket{\psi|U|\psi}\right)
\end{equation} 

The Hadamard test can be utilized to serve our purpose by defining $U=E(y)^\dagger E(x)$ and $\ket{\psi} = \ket{0}$, which simplifies $Re\left(\braket{\psi|U|\psi}\right)$ to $\braket{0|E(y)^\dagger E(x)|0}$. Similar to the Swap Test, the Hadamard analogue faces the potential challenge of running into higher circuit depth due to the presence of the $CU$ gate. The Hadamard-Overlap test described in \cite{overlap} partially bypasses the problem of controlled unitaries at the cost of using almost double the number of qubits in the circuit.

\subsubsection{Simple Overlap Calculation} \label{sec:simple}
\begin{figure}[h]
\[ \Qcircuit @C=2em @R=1em @!R{
	\lstick{\ket{0}} & \multigate{1}{E(x)} & \multigate{1}{E(y)^\dagger} & \multigate{1}{\ket{0_n}\bra{0_n}} & \cw \\
	\lstick{\ket{0}} & \ghost{E(x)} & \ghost{E(y)^\dagger} & \ghost{\ket{00}\bra{00}}
& } \]
\caption{Simple circuit to compute similarity between two quantum states.}
\label{fig:overlap}
\end{figure}
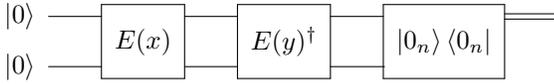

The circuit in Fig. \ref{fig:overlap} show perhaps the simplest and most economical algorithm that can be used to calculate the similarity of two quantum statevectors. The method estimates $\braket{y|x} = \braket{0|E(y)^\dagger E(x)|0}$  by consecutively applying the encoding unitaries $E(x)$ and $E(y)^\dagger$ to the same set of qubits, without the application being controlled by another qubit. The requisite value is then extracted by finding the probability of measuring all the qubits in the state $0$ by running the circuit for multiple shots. Alternatively, the expectation value of the zero-projector, $\ket{0}_n\bra{0}_n$, also provides equivalent results. Owing to the shallow depth and low number of qubits required, this algorithm emerges as the most conducive choice to find $D_E^2$ in the implementation of $q$-means algorithm.  

The depicted circuits of all the algorithms described for distance estimation or similarity evaluation must be run for multiple shots to glean the requisite information from them. Naturally, higher the number of shots used, better are the results obtained. This acts as a NISQ-aligned alternative to the use of amplitude estimation \cite{QAE}, or variants thereof \cite{IQAE, MLQAE}, to extract the same information from the circuits, which comes at the cost of higher number of qubits required and much higher circuit depth, both of which are a bane of NISQ processors.

\subsection{Image Segmentation}\label{sec:imgsegment}
Image Segmentation is typically used to locate objects and boundaries (lines, curves, etc.) in images. More precisely, image segmentation is the process of assigning a label to every pixel in an image such that pixels with the same label share certain characteristics \cite{morioh}. A lot of research has been done in the area of image segmentation using clustering. There are a plethora of methods and one of the popular ones is to use the $k$-means clustering algorithm. Although this algorithm was not originally developed specifically for image processing, it has been adopted by the computer vision community and is used up to these days \cite{k-means}. The $k$-means algorithm requires the a priori knowledge of the number of clusters ($k$) into which the image pixels should be grouped. Each pixel of the image is repeatedly and iteratively assigned to the cluster whose centroid is closest to the pixel.

The $q$-means algorithm has been leveraged for the purpose of image clustering as follows: the images were converted to grayscale, as the pixel-intensity was found to be the qualifying feature to identify whether it belongs to a crack-region or not. The number of available qubits being limited in the NISQ-era, the images were also downscaled from $227\times 227$ to a resolution of $50\times 50$, post which a $5\times 5$ Gaussian blurring filter was applied to get rid of noise and somewhat irrelevant high frequency components from the images. The new resolution was chosen arbitrarily and was visually found to retain the discriminating features between the cracks and background regions. The complexity of the images in the data set being low, the number of clusters has been set to two, i.e., $k=2$; and the cluster-centroids are specified by $8$-bit integers. Henceforth, the intensities of the pixels, which range between $[0,255]$, have been treated as single-featured vectors of a data set.

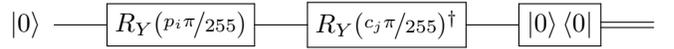
\begin{figure}[h]
	\[ \Qcircuit @C=2em @R=1em @!R{
		\lstick{\ket{0}} & \gate{R_Y(\nicefrac{p_i\pi}{255})} & \gate{R_Y(\nicefrac{c_j\pi}{255})^\dagger} & \gate{\ket{0}\bra{0}} & \cw } \]
	\caption{Circuit to find distance of $i^{th}$ pixel-intensity from $j^{th}$ cluster-centroid.}
	\label{fig:perpixel}
\end{figure}

The simple overlap calculation method, described in Sec \ref{sec:simple} has been used to calculate the distances of the data vectors from the cluster-centroids. Angle embedding using $R_Y$ gates has been used to encode the intensity and centroid information, post appropriate scaling, into a single qubit per pixel to preserve the simplicity of the procedure. If intensity of the $i^{th}$ pixel is $p_i$, and centroid of the $j^{th}$ cluster is $c_j$, then $E(p_i) = R_Y(\nicefrac{p_i\pi}{255})$ and $E(c_j) = R_Y(\nicefrac{c_j\pi}{255})$, resulting in a per-pixel circuit given in Fig. \ref{fig:perpixel}. Why this approach works for our distance calculations and how it compares with the classical calculations are captured in Appendix \ref{sec:app}; the results reinforce the usability of the suggested distance.

\begin{figure}
	\centering
	\begin{subfigure}{0.24\textwidth}
		\centering
		\includegraphics[width=\textwidth]{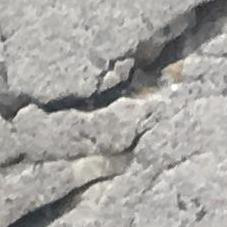}
		\caption[Network2]%
		{{\small Original image from data set containing cracks}}    
		\label{fig:orig}
	\end{subfigure}
	\begin{subfigure}{0.24\textwidth}  
		\centering 
		\includegraphics[width=\textwidth]{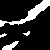}
		\caption[]%
		{{\small Result of $q$-means on downscaled image}}    
	\end{subfigure}
	\begin{subfigure}{0.24\textwidth}   
		\centering 
		\includegraphics[width=\textwidth]{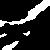}
		\caption[]%
		{{\small Result of $k$-means on downscaled image}}    
	\end{subfigure}
	\begin{subfigure}{0.24\textwidth}   
		\centering 
		\includegraphics[width=\textwidth]{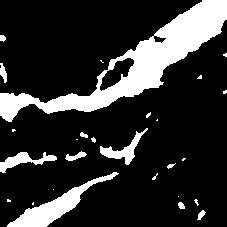}
		\caption[]%
		{{\small Result of $q$-means on original image}}    
	\end{subfigure}
	\caption[ The average and standard deviation of critical parameters ]
	{\small Results of $k$-means and $q$-means image segmentation on $50\times 50$ and $227\times 227$ versions of an image with cracks.} 
	\label{fig:segmentationresults}
	\vspace{-0.25cm}
\end{figure}

The aforesaid process for image segmentation through $q$-means was carried out by iterating over each of the $2500$ pixels in the downscaled image and over the two cluster-centroids. This implied the use of $2500\times 2 = 5000$ qubits to find the Euclidean-equivalent distances per iteration of the $q$-means algorithm. Presently, this is way beyond the number of qubits that are available at our disposal through any vendor of quantum processors. Also of note is the fact that most popular vendors that provide the use of their quantum processors through the cloud, charge per circuit and the number of shots that the circuits are executed for, and not the overall width of the circuit. Thus, the most economical option is to use as many qubits per circuit as possible, a proposition which is not always advisable due to the reduced connectivity of the qubits in a processor which may lead to great circuit depths due to repeated swapping of the qubits. On fully-connected devices like IonQ's \cite{IonQprocessor}, however, this is easily amenable. Since the proposed overlap-calculation circuit does not use entanglement (which does not provide a quantum advantage, but again, this is a place-holder approach until such time a QRAM is available), the qubits are batched into groups of $n$ for pixel-centroid distance estimation, where $n$ is the number of qubits that are available. The results of $q$-means image segmentation on AWS Braket Simulator of an image with cracks (original resolution and downscaled) and comparison with classical $k$-means based segmentation have been portrayed in Fig. \ref{fig:segmentationresults}. It must be noted that the simulation times for the $q$-means algorithm acting on downscaled images ranged between $60-90$ seconds per image.

\subsection{Quantum Classifier}
Quantum machine learning has the potential for broad industrial applications, and the development of quantum algorithms for improving the performance of neural networks is of particular interest given the central role they play in machine learning today \cite{allock}. 

Feedforward neural networks play a key role in machine learning, with applications ranging from computer vision and speech recognition to data compression and recommendation systems. In a supervised learning scenario, a network is trained to recognize a set of labelled data by learning a hierarchy of features that together capture the defining characteristics of each label. Once trained, the network can then be used to recognize unlabelled (test) data. 

Quantum algorithms, i.e., algorithms that can be executed on a quantum computer, have been investigated since the 1980s, and have recently received increasing interest all around the world. One of the main applications for quantum computing is the development of new algorithms for machine learning \cite{vqc2}. Variational quantum algorithms (VQAs) use classical optimizers to train parametrized quantum circuits and represent an advancement in quantum computing running on NISQ computers. Variational algorithms can be the basis for numerous applications, including the design of a quantum classifier. 

In our proposed pipeline, we designed Variational Quantum Classifiers (VQC) to carry out the requisite classification at two places (see Fig. \ref{fig:detailed}); the first classifier learns how to classify the whole input image to pass for crack detection or not. These images are segmented they are in turn tagged to cracks or not by the second classifier.  A VQC itself consists of three stages: 
\begin{enumerate}
\item State Preparation, which also encodes the classical data (feature vector) for further quantum processing, \item A model circuit or Ansatz with optimizable parameters, \item A measurement stage. 
\end{enumerate}
VQC generally consists of a number of one-qubit gates: the Hadamard gate $H$, and rotational gates; plus some entangling gates: the controlled-NOT and the controlled-Z. Very specifically, in our work we used angle encoding for the classical features and incorporated strongly entangling layers \cite{sel} in the Ansatz. The classical counterpart, completing the hybrid nature of VQC utilize the measurement outputs to evaluate a appropriately chosen cost function and further minimizing it either through the use of gradients or invoking non-gradient approaches (see also \cite{vqc3}).  

Some of the advantages of Variational Quantum Classifiers include - A VQC can outperform a classical model using far less free parameters and, thus, being more efficient. Further, a complex classification task requires deeper quantum circuits, which nevertheless grow at a slower pace than the number of
neurons needed in a Classical Neural Network for the same task \cite{vqc3}. Also, in some cases, VQCs can perform similar to (classical) classifiers with a lesser number of training samples.

\subsubsection{Data Dimensionality Reduction}
Large data sets are increasingly widespread in many disciplines. In order to interpret such data sets, methods are required to drastically reduce their dimensionality in an interpretable way, such that most of the information in the data is preserved. Many techniques have been developed for this purpose, but principal component analysis (PCA) is one of the oldest and most widely used. Its idea is simple—reduce the dimensionality of a data set, while preserving as much `variability' (i.e.statistical information) as possible \cite{pca1}. The latter translates into finding new variables that are linear functions of those in the original data set, that successively maximize variance and that are uncorrelated with each other. Finding such new variables, the principal components (PCs), reduces to solving an eigenvalue/eigenvector problem \cite{pca1}.

As depicted in our pipeline, PCA is used as a dimensionality reducing preprocessing stage, whose output are used as “classical” features input to the VQC. The rationale is rather simple: since the dimension of the data corresponds to the number of qubits required in order to encode the data, and further, we can handle limited number of qubits both in the simulators and the NISQ computers, reduction in dimension is inevitable.

Needless to say, we have considered classical PCA for our purpose. (Classical) PCA involves a computational cost of $O(N^2)$, where, $N$ is the size of the data vector, and the $n\times N$ covariance matrix of data is subjected to eigen-decomposition. It is useful to note that a Quantum PCA exists which has the computational cost of $O((\log N)^2)$ \cite{pca2}. In future, when there is a progress in quantum computing hardware, including a form of QRAM, one can utilize the Quantum PCA for further speed up of the proposed pipeline execution.

\subsubsection{Encoding}
\begin{figure}[h]
	\[ \Qcircuit @C=2em @R=1em @!R{
		\lstick{\ket{0}} & \gate{H} & \gate{R_Z(x_0)} & \qw \\
		\lstick{\ket{0}} & \gate{H} & \gate{R_Z(x_1)} & \qw \\
		\lstick{\ket{0}} & \gate{H} & \gate{R_Z(x_2)} & \qw \\
		\lstick{\ket{0}} & \gate{H} & \gate{R_Z(x_3)} & \qw 
} \]
	\caption{$R_Z$ encoding that encodes the features represented by $x_i$s into the phases of quantum states.}
	\label{fig:rzencoding}
\end{figure}
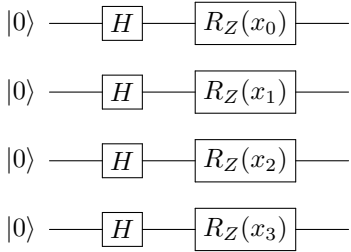

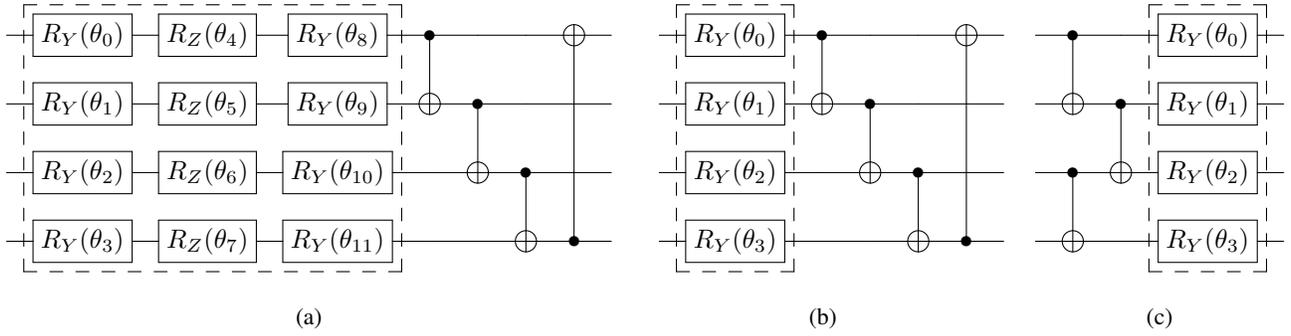
\begin{figure*}[t]
	\begin{subfigure}{0.5\textwidth}
		\centering
		\[ \Qcircuit @C=1em @R=1em @!R{
			\lstick{} & \gate{R_Y(\theta_0)} & \gate{R_Z(\theta_4)} & \gate{R_Y(\theta_8)} &	\ctrl{1} &	\qw & 		\qw &		\targ &		\qw\\
			\lstick{} & \gate{R_Y(\theta_1)} & \gate{R_Z(\theta_5)} & \gate{R_Y(\theta_9)} &	\targ &		\ctrl{1} & 	\qw &		\qw &		\qw\\
			\lstick{} & \gate{R_Y(\theta_2)} & \gate{R_Z(\theta_6)} & \gate{R_Y(\theta_{10})} &	\qw &		\targ & 	\ctrl{1} &	\qw &		\qw\\
			\lstick{} & \gate{R_Y(\theta_3)} & \gate{R_Z(\theta_7)} & \gate{R_Y(\theta_{11})} &	\qw &		\qw &		\targ &		\ctrl{-3} &	\qw \gategroup{1}{2}{4}{4}{.7em}{--}
		} \]
		\caption{}
		\label{fig:stronglyentangling}
	\end{subfigure}
	\begin{subfigure}{0.24\textwidth}
		\centering
		\[ \Qcircuit @C=1em @R=1em @!R{
			\lstick{} & \gate{R_Y(\theta_0)} &	\ctrl{1} &	\qw & 		\qw &		\targ &		\qw\\
			\lstick{} & \gate{R_Y(\theta_1)} &	\targ &		\ctrl{1} & 	\qw &		\qw &		\qw\\
			\lstick{} & \gate{R_Y(\theta_2)} &	\qw &		\targ & 	\ctrl{1} &	\qw &		\qw\\
			\lstick{} & \gate{R_Y(\theta_3)} &	\qw &		\qw &		\targ &		\ctrl{-3} &	\qw \gategroup{1}{2}{4}{2}{.7em}{--}
		} \]
		\caption{}
		\label{fig:basicentangling}
	\end{subfigure}
	\begin{subfigure}{0.24\textwidth}
		\centering
		\[ \Qcircuit @C=1em @R=1em @!R{
			\lstick{} & \ctrl{1} & \qw & \gate{R_Y(\theta_0)}&		\qw\\
			\lstick{} & \targ & \ctrl{1} & \gate{R_Y(\theta_1)} &		\qw\\
			\lstick{} & \ctrl{1} & \targ & \gate{R_Y(\theta_2)}&		\qw\\
			\lstick{} & \targ & \qw & \gate{R_Y(\theta_3)} &	\qw \gategroup{1}{4}{4}{4}{.7em}{--}
		} \]
		\caption{}
		\label{fig:pennylaneansatz}
	\end{subfigure}
	\caption{(a) First layer of the Strongly Entangling Layers ansatz. The subsequent layers feature parametrized $R_YR_ZR_Y$ gates followed by $CNOT$ gates with skip connections, with the skip-level depending on the layer-number. (b) First layer of the Basic Entangling Layers ansatz. Similar to the Strongly Entangling ansatz, the subsequent layers feature parametrized $R_Y$ gates followed by $CNOT$ gates with skip connections, with the skip-level depending on the layer-number. (c) A single layer of the ansatz used in the Pennylane tutorial. The dashed boxes include rotational gates with optimizable parameters.}
	\label{fig:ansaetze}
\end{figure*}

Of the myriad techniques available for classical-to-quantum data conversion and embedding them into circuits, the most simple method, i.e., angle-encoding was selected. More specifically, $H$ gates followed by $R_Z$ gates with the features as parameters have been used, which incorporates the said features as phases into the quantum states. The combination of $H$ and $R_Z$, as in Fig. \ref{fig:rzencoding}, was found to work slightly better than the other rotation gates available for angle-encoding, namely the $R_X$ and $R_Y$ gates. Basis-encoding would not have been conducive for embedding image features as it caters only to binary data. Amplitude encoding, on the other hand, would have been suitable for the task and advantageous due to the lower number of qubits required, but was neglected in favour of angle-encoding due to the overall difficulty and increased circuit-depth of the process.

\subsubsection{Ansatzes}

An ansatz is a block of gates with optimizable parameters that selects a basis of measurement appropriate for the problem at hand by iterating over the training inputs for multiple epochs and trying to match the post-processed measurements with the ground-truth values corresponding to each example. Similar to the bias-variance conundrum in classical machine and deep learning, an ansatz may provide a better expressibility or may choose to increase the entanglement capability of the qubits in a circuit \cite{ene}. Similarly, an ansatz may be constructed to be efficiently implementable on processors with a certain kind of qubit-connectivity, or hardware-agnostic versions of it can be created with just the complexity of the problem in mind.

To address the problem of image and segment classification, the three ansatzes (only single layers of which are displayed in Fig. \ref{fig:ansaetze}) have been utilized. The ansatz in Fig. \ref{fig:stronglyentangling}, has the highest circuit depth of the three and uses three parametrized gates per qubit, per layer of the ansatz \cite{sel}. It is widely known that any single-qubit unitary can be expressed as a combination of $R_YR_ZR_Y$ gates, up to a global phase. The rotation gates are followed by multiple $CNOT$ gates, the target qubit of each gate is determined using the relation $t = (c+l+1) \; mod \; n$, where $c$ and $t$ are the control and target qubit numbers, $l$ is the ansatz-layer being constructed and $n$ is the number of qubits the ansatz operates on. An $L$-layer ansatz avails $3nL$ optimizable parameters.

The ansatz in Fig. \ref{fig:basicentangling} is similar in nature, but uses only an $R_Y$ gate per qubit, which reduces the number of parameters to just $nL$. In contrast to the previous two, the third ansatz in Fig. \ref{fig:pennylaneansatz} (employed in \cite{pennylane}) is constructed using a fixed topology of entangling gates. Since the $CNOT$ gates appear before the rotation ones in each layer, this ansatz is incompatible with phase-encoding through $R_Z$ gates, and hence has been used in conjunction with angle-encoding in the form of $R_Y$ rotations, instead.

\subsubsection{Measurement}
Post the application of the encoding and the ansatz layers, the quantum information must be brought back to the classical world through the process of measurement. In an otherwise linear manipulation of data by the various gates in the circuit, measurement provides the requisite non-linearity to the VQC \cite{Schuld}. Various measurement and post-processing techniques have been to glean information from the circuits - using the parity of qubit-readout \cite{QSVM}, measuring the expectation values in a certain basis and passing them along to another classical neural network for inference \cite{Schuld}, etc., are among the popular approaches.

The expectation value with respect to the Pauli-$Z$ operator of the first qubit, which lies in the range of $[-1,1]$, has been used in this work as the feature that segregates each data point into either of the two classes in the binary-classification task. If the measured value is beyond a certain threshold, that is determined by the classical optimizer, then the data point is assigned to a particular class; otherwise, the other class is chosen for the given input.

\subsubsection{Classical Optimizer}
To accrue some benefit out of Noisy Intermediate-Scale Quantum computers that are plagued with limited number of qubits, noisy and decoherence, the use of hybrid quantum algorithms has been suggested. In such algorithms, certain tasks that are amenable on quantum are offloaded to quantum processors, and classical and quantum counterparts work in tandem to solve a larger problem. In variational quantum algorithms, more often than not, the optimization of parameters is entrusted to the classical processors.

Within the scope of this work, the gradient-free $COBYLA$ optimizer from the Scipy python package has been employed. The calculation of quantum gradients using methods like the parameter-shift rule \cite{psr} was avoided to keep the number of shots and circuit simulation times in check. To the best of the authors' knowledge, AWS Braket SDK does not provide an interface to TensorFlow, PyTorch, or any of the popular deep learning libraries. This hinders the creation of computational graphs from the circuits, rendering the advanced optimizers from the aforesaid libraries unusable.
The classical optimizer was tasked with minimizing the Mean Squared Error (MSE) between the ground-truth values and the predictions from the VQC.

\subsection{Image Classification}
As illustrated in Sec. \ref{sec:overall}, a classifier is run on each image to qualify whether it contains a crack or not. The rest of the inference pipeline is executed only if it is determined that the image does indeed have crack(s) in it. Over the following subsections, the training process for such a pipeline has been described in detail.

\subsubsection{Data set and Preprocessing} 
The data set described in Sec. \ref{sec:dataset} is a balanced one, i.e., it contains an even $20,000$ images for each category. However, in an industrial setting, anomalous data is more difficult to come by. With an exterme dearth of data for a particular task, the training of efficient classifiers is a precarious affair. This is a peril for which quantum classifiers seem to provide an anticipation of a remedy \cite{hep}. A similar impediment has been simulated by selecting a subset of the data with a modest imbalance. $1000$ images were randomly chosen from the overall data set, with the precondition that only $25\%$ of which would be images that have cracks in them. The dimensionality-reduced VGG16 features of the $1000$ images were further preprocessed by scaling them to the range of $[0, \pi]$. $25\%$ of these reduced and scaled feature vectors were set aside for testing, and of the remaining $75\%$, a further $25\%$ were utilized only for the purpose of validation. As a result, the training was performed on only about $563$ images.

\subsubsection{Training}\label{sec:imgtraining}
\begin{figure}[h]
	\centering
	\begin{subfigure}{0.24\textwidth}
		\centering
		\includegraphics[width=\textwidth]{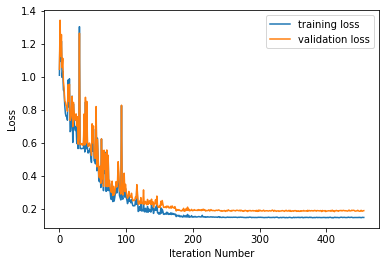}
		\caption[Network2]%
		{{\small Losses for circuit with the ansatz in Fig. \ref{fig:stronglyentangling}.}}    
	\end{subfigure}
	\begin{subfigure}{0.24\textwidth}  
		\centering 
		\includegraphics[width=\textwidth]{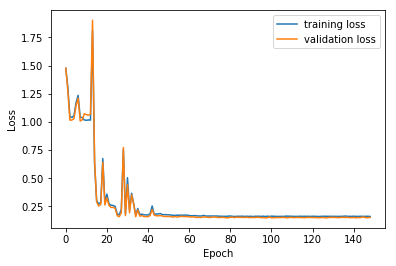}
		\caption[]%
		{{\small Typical losses for circuits with ansatzes from Figs. \ref{fig:basicentangling} and \ref{fig:pennylaneansatz}.}}    
	\end{subfigure}
	\caption[ The average and standard deviation of critical parameters ]
	{\small Plot of training and validation loss per iteration encountered while training the image classification circuit.} 
	\label{fig:losses}
	\vspace{-0.25cm}
\end{figure}

The variational quantum model for image classification was trained with the data set described above using phase-encoding, the three Ansatzes in Fig. \ref{fig:ansaetze} individually with $n=4$, $L=3$, $1000$ shots per circuit, and the MSE loss was reduced using the COBYLA optimizer. Without gradient-calculation and with $36$, $12$ and $12$ parameters, each, the loss from the circuits with the three ansatzes converged in about $90$, $15$ and $15$ minutes, respectively. In keeping with the number of parameters to be optimized, the first circuit ran for about $450$ epochs before convergence, while the other two required only about $150$ iterations, each. A typical plot of the training and validation losses against the corresponding iteration numbers can be found in Fig. \ref{fig:losses}. Again, in the absence of gradient-based optimizers, the optimizer blindly bounces around in the loss landscape, taking larger steps initially. Gradually, it converges to a value, but continues to run further iterations before finally terminating the optimization procedure. Having fewer parameters to optimize significantly reduces the jitter in loss, along with the number of epochs and time required to converge. 

The testing accuracy metrics for the VQC models with the three different types of ansatzes for the test data set have been reported in table \ref{tab:accuracies}. The number of misclassifications although being low, were mostly false negatives, which are detrimental because the subsequent pipeline does not get triggered for those images. This is an artefact of the artificial imbalance introduced in the data set, the effect of which can be offset through the use of more sophisticated loss functions, such as \textbf{Focal Loss}, that are more sensitive to imbalanced data.

\subsection{Segment Classification}
The images that are adjudges by the image classification model to contain cracks are passed to the segmentation algorithm in Sec. \ref{sec:imgsegment}. The segmented images exemplified in Fig. \ref{fig:segmentationresults} have multiple regions that may either be cracks, or may belong to the other aberrations discussed about in Sec \ref{sec:dataset} and shown in Fig. \ref{fig:negative}. Each such region is passed along to another variational quantum classification model that has been dubbed segment classifier.

\subsubsection{Data set and Preprocessing}

\begin{figure}[h]
	\centering
	\begin{subfigure}{0.24\textwidth}
		\centering
		\includegraphics[width=\textwidth]{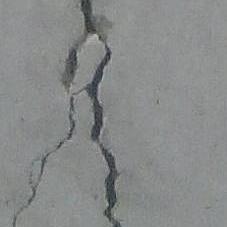}
		\caption[Network2]%
		{{\small Original image}}    
	\end{subfigure}
	\begin{subfigure}{0.24\textwidth}
		\centering
		\includegraphics[width=\textwidth]{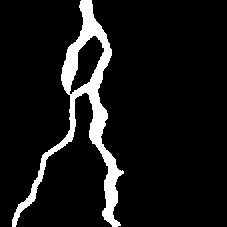}
		\caption[Network2]%
		{{\small Manual annotation of cracks}}    
	\end{subfigure}
	\begin{subfigure}{0.24\textwidth}
		\centering
		\includegraphics[width=\textwidth]{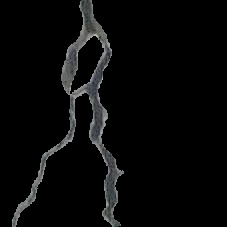}
		\caption[Network2]%
		{{\small Image with only region 1}}    
	\end{subfigure}
	\begin{subfigure}{0.24\textwidth}  
		\centering 
		\includegraphics[width=\textwidth]{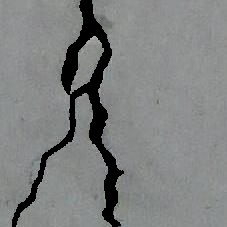}
		\caption[]%
		{{\small Image with only region 2}}    
	\end{subfigure}
	\caption[ The average and standard deviation of critical parameters ]
	{\small Figures show the original image considered, the masks created manually for the crack-regions, a blank image containing only the crack-region, and another blank image with only the background.} 
	\label{fig:window}
	\vspace{-0.25cm}
\end{figure}

To create the data set to train the segment classifier, a corpus of $150$ images with cracks was created through random selection and the cracked regions were manually annotated. The different regions of the annotated images were identified through contour detection, and each region (including the background) were extracted out separately, as shown in Fig. \ref{fig:window}. The VGG16 features of all such region-images was obtained and stored along with their labels to identify whether the regions correspond to cracks or not. Further preprocessing in the form of dimensionality-reduction, scaling, and splitting for training, validation and testing were carried out as described before, considering the same percentages.

\subsubsection{Training}
Training for the segment classifier was carried out with the same circuits and methods described in Sec. \ref{sec:imgtraining}. The segment classifiers  with the ansatzes in Fig. \ref{fig:ansaetze}, being trained on a more balanced data set, and on a considerably lesser number of examples, took only about $20$, $5$ and $5$ minutes, each. The plots of loss vs epochs also were similar to those in Fig. \ref{fig:losses} for the three ansaetze. The segment classification accuracy metrics for the three trained models on the test data set are captured in table \ref{tab:accuracies}.

\subsubsection{Post-processing}
\begin{figure}[h]
	\centering
	\begin{subfigure}[b]{0.24\textwidth}
		\centering
		\includegraphics[scale=0.5]{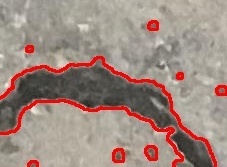}
		\caption[Network2]%
		{{\small }}    
		\label{fig:misclass}
	\end{subfigure}
	\begin{subfigure}[b]{0.24\textwidth}  
		\centering 
		\includegraphics[width=\textwidth]{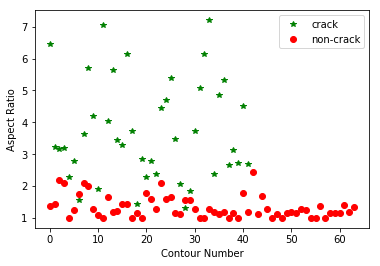}
		\caption[]%
		{{\small }}    
		\label{fig:ratios}
	\end{subfigure}
	\caption[ The average and standard deviation of critical parameters ]
	{\small (a) Result of segment classification on an image with false-positives. (b) Plot of aspect ratios of oriented bounding boxes for crack and non-crack regions.} 
	\label{fig:postprocessing}
	\vspace{-0.25cm}
\end{figure}

Unlike the image classifier, the segment classifier is plagued with the problem of false-positives. Many regions of the images which are not cracks, get misidentified as such. Fig. \ref{fig:misclass} shows an image where although the crack is segmented properly, many additional regions are also denoted as cracks by the classifier. The difficulty may be remedied by enclosing the regions identified as cracks in oriented bounding boxes and finding the aspect ratio of the latter. A plot of the aspect ratios for crack and non-crack regions can be found in Fig. \ref{fig:ratios}. The fairly-separable distribution of ratio-values in the plot suggests that a simple classical or quantum classifier can be trained to filter out the non-crack regions getting identified as cracks, though at the cost of increased false-negatives.

\section{Classical Approaches}\label{sec:classical}
To benchmark the performance of the quantum approaches discussed to solve the problem of crack-detection, it was compared against two classical solutions. One of the solutions was devised by replacing the quantum blocks (i.e., the quantum classifier and the $q$-means algorithm) in the pipeline detailed in Sec. \ref{sec:detailed} with their direct classical counterparts. The $4$-dimensional PCA-reduced VGG16 features from an image were passed to a simple fully-connected layer to get binary, categorical output, and the binary cross entropy loss was minimised using the $Adam$ optimizer. The other approach involved trials with a home-brewed deep learning model called SCNet  \cite{scnet} built specifically for surface crack detection. The models were trained, validated and tested against the same set of data as their quantum analogues, although for SCNet, the images had to be augmented. The performance metrics of the classical approaches have also been briefed in table \ref{tab:accuracies}.

\section{Overall Results and Accuracy}\label{sec:results}
\begin{figure}[h]
	\centering
	\begin{subfigure}{0.24\textwidth}
		\centering
		\includegraphics[width=\textwidth]{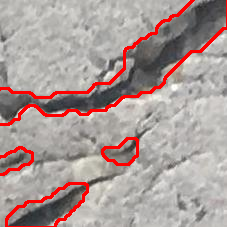}
		\caption[Network2]%
		{{\small }}    
	\end{subfigure}
	\begin{subfigure}{0.24\textwidth}
		\centering
		\includegraphics[width=\textwidth]{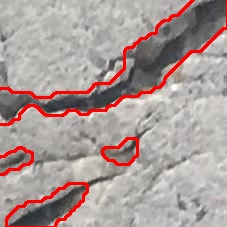}
		\caption[Network2]%
		{{\small }}    
	\end{subfigure}
	\begin{subfigure}{0.24\textwidth}
		\centering
		\includegraphics[width=\textwidth]{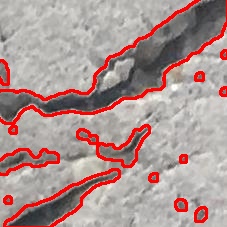}
		\caption[Network2]%
		{{\small }}    
	\end{subfigure}
	\begin{subfigure}{0.24\textwidth}  
		\centering 
		\includegraphics[width=\textwidth]{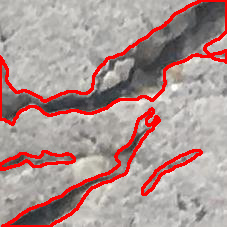}
		\caption[]%
		{{\small }}    
	\end{subfigure}
	\caption[ The average and standard deviation of critical parameters ]
	{\small (a) shows the results from a classical counterpart of the same pipeline used for quantum. (b) and (c) show the results obtained from the quantum models with the Strongly Entangling Layers ansatz, without any post-processing. For (a) and (b), the original image was downscaled to speed up the process of image segmentation. (d) portrays the results from the in-house classical crack-segmentation model, SCNet.} 
	\label{fig:results}
	\vspace{-0.25cm}
\end{figure}

A further $150$ images were annotated to test the accuracy of the end-to-end pipeline, which was quantified using Intersection over Union (IoU) of the segmented cracks with the ground-truth annotations. Along with the accuracy value of various classical and quantum classification models (without downscaling of images), the respective IoU scores of the overall pipeline have also been included in table \ref{tab:accuracies}. The segmentation results from some of the approaches for the image in Fig. \ref{fig:orig} are shown in Fig. \ref{fig:results}.

\begin{table}[h]
	\renewcommand\arraystretch{1.5}
	\centering
	\resizebox{0.495\textwidth}{!}{
	\begin{tabular}{| c | c | c | c |}
		\hline
		 & \textbf{Image Classification} & \textbf{Segmentation Classification} & \textbf{IoU} \\
		\hline
		\textbf{Ansatz in Fig. \ref{fig:pennylaneansatz}} & 97.6 & 98.7 & 72.76 \\ 
		\hline
		\textbf{Basic Entangling Layers} & 94.8 & 97.4 & 73.14 \\
		\hline
		\textbf{Strongly Entangling Layers} & 99.2 & 98.7 & 73.33 \\
		\hline
		\textbf{Classical pipeline} & 100 & 100 & 75.21 \\
		\hline
		\textbf{SCNet} & N/A & N/A & 91.36 \\
		\hline
	\end{tabular}
}
\caption{Accuracy values for image and segment classification along with IoU scores for various methods.}
\label{tab:accuracies}
\end{table}

The low accuracy values reported for the quantum models may be due to the drastic reduction in dimensionality to fit the contemporary quantum simulators and hardware. Of course, further research is also possible in terms of data encoding, etc. It is also noteworthy that the explored quantum models use around $12-36$ parameters and get trained in a fraction of the time compared to SCNet which depends on about \textit{30 million} trainable parameters!

\section{Future Work}\label{sec:future}
The work carried out for this paper opened up many threads for explorations. As mentioned in the previous section, one immediate thing to consider is arrive at other suitable data encoding schemes, say, dense angle encoding which facilitates encoding two features per qubit \cite{encoding}. Since we have carried out only simulations in AWS, running on different hardware is another task, including comparative performance and benchmarking. 
	
Within the context of classification and segmentation, extending the research to multi-class scenario is another useful and challenging direction. In this paper, we considered a data set with a number of examples; in case, if only few images are available, generating a sized number  of images \cite{imggen} for training is a practical and an important issue to address.

Going tangentially off, since the Deep Learning based technique of \cite{scnet} performed well for segmentation, ways to ``quantize" it can be considered as well.

\section{Quantum Software Engineering}\label{sec:qse}
The increasing complexity involved in the solutions developed using quantum computing has mandated the formulation and formalisation of a Quantum Software Development Life Cycle \cite{qdlc}, which is formed by the interweaving of quantum, machine learning and software engineering life cycles. None of the components can be considered independently, and a holistic view of the system is necessary for proper functioning. Such a life cycle involves all of the machine learning and software engineering life cycle steps, in addition to certain-quantum specific parts such as selection of data-encoding strategies, quantum-models etc., among others.  

While experimenting with Quantum Machine Learning models (which are invariably a mix of classical and quantum techniques), keeping tabs on all the classical options is a daunting task in itself, which is augmented by the plethora of quantum options, rendering manual tracking of the models, parameters, metrics, performance indicators, version-tracking, and the permutations and combinations thereof, an unsurmountable feat. Further, quantum computing being at a nascent stage, is nowhere close to being productionized. As a result, most practitioners train, validate and test their developed models within the same file or jupyter notebook. The possibility of saving quantum models for future inferencing and use is rarely considered or discussed. 

Trained QML models which consist mostly of the optimized weights, along with the helper functions, in the form of objects, can be easily stored in the form of \textit{pickle} or \textit{dill} files if the quantum SDK being used is python-based. Additionally, to address non-functional aspects of the problem like performance and experiment tracking, reproducibility, orchestration, distributed execution etc., Cognitive Model Management (CMM) was used, which is a framework, intended to help the user manage the lifecycle of the quantum and (or) classical models. CMM provides modular API-based architecture to obtain maximum agility and scalability along with orchestration facility where the user can orchestrate workflows from tasks and execute them in a distributed and clustered environment. Finally, when the models were trained, another self-developed tool called OpSense  was leveraged to track the experiments and deploy the model that had the best metrics.

\section*{Conclusion}
Set out to solve the image classification and segmentation, which can be of great use across domains, the paper brought out a generic hybrid quantum-classical pipeline. Each quantum block was worked out in a systematic manner and the entire pipeline was assessed for its functionality and performance by considering Kaggle Surface Crack Detection data. Comparison with other fully classical pipelines and the associated metrics were captured. In order to systematically handle the software-engineering aspects, we also adopted the Cognitive Model Management framework as well as another tool called OpSense, and successfully put into action the entire work flow involved.

\section*{Acknowledgement}
The authors would like to sincerely thank Mr. Anil Sharma, Head of TCS Incubation, and Dr. J. Gubbi, Senior Scientist, TCS Research, for their constructive feedback and support, without which this work would not have been possible.

\begin{appendices}
\section{Classical vs Quantum Distance Calculation}\label{sec:app}
\begin{figure}[h]
	\centering
	\includegraphics[width=.3\textwidth]{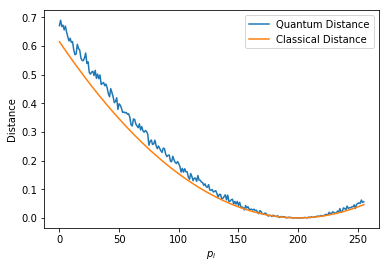}
	\caption{Comparison of quantum and classical distance estimations for $c_j=200$ and $p_i$ ranging from $0$ to $255$, at unit intervals.}
	\label{fig:distplot}
\end{figure}
\begin{figure}[t]
	\centering
	\begin{subfigure}[b]{0.24\textwidth}
		\centering
		\includegraphics[width=\textwidth]{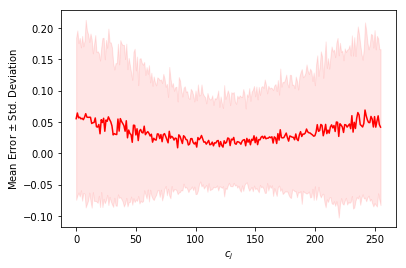}
		\caption[Network2]%
		{{\small 10 shots}}    
		\label{fig:mean and std of net14}
	\end{subfigure}
	\hfill
	\begin{subfigure}[b]{0.24\textwidth}  
		\centering 
		\includegraphics[width=\textwidth]{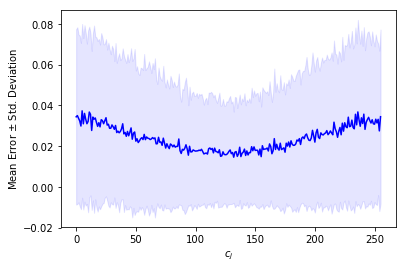}
		\caption[]%
		{{\small 100 shots}}    
		\label{fig:mean and std of net24}
	\end{subfigure}
	\vskip\baselineskip
	\begin{subfigure}[b]{0.24\textwidth}   
		\centering 
		\includegraphics[width=\textwidth]{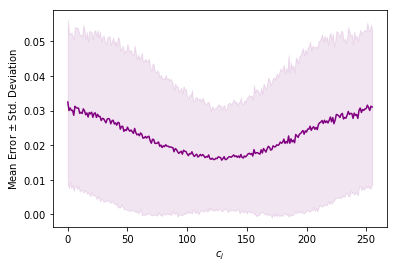}
		\caption[]%
		{{\small 1000 shots}}    
		\label{fig:mean and std of net34}
	\end{subfigure}
	\hfill
	\begin{subfigure}[b]{0.24\textwidth}   
		\centering 
		\includegraphics[width=\textwidth]{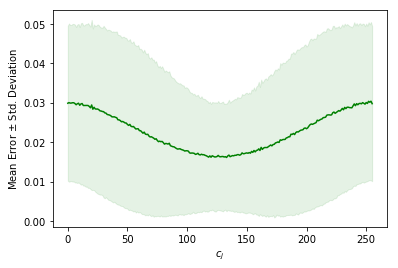}
		\caption[]%
		{{\small 10000 shots}}    
		\label{fig:mean and std of net44}
	\end{subfigure}
	\caption[ The average and standard deviation of critical parameters ]
	{\small The average and standard deviation of errors in Euclidean distance calculation for various number of shots, with various values of $c_j$ on the $x$-axis.} 
	\label{fig:shotcomparison}
\end{figure}

The suitability of the distance-calculation approach can be verified by the plot in Fig. \ref{fig:distplot}, which compares the distances estimated using simulated quantum circuits against scaled classical distances $\left(\nicefrac{p_i-c_j}{255}\right)^2$. The number of shots considered for this validation exercise was $1000$ per circuit. A comparative study of the results against the number of shots can be found in the plots of Fig. \ref{fig:shotcomparison}, where both $p_i$ and $c_j$ were varied from $0$ to $255$ and the average error per value of $c_j$ has been shown as the baseline, surrounded by $\pm$ standard deviation of errors. As expected, the curves become smoother and error reduces as the number of shots is increased, due to better estimation of expectation values.

\end{appendices}

\end{document}